# Femtosecond laser induced structural dynamics and melting of Cu (111) single crystal. An ultrafast time-resolved x-ray diffraction study.

Runze Li[1], Omar A. Ashour[1], Jie Chen[2], H.E. Elsayed-Ali[3], and Peter M. Rentzepis[1,*]

[1]Department of Electrical and Computer Engineering, Texas A&M University, College Station, TX 77843, USA

[2]Key Laboratory for Laser Plasmas (Ministry of Education), Department of Physics and Astronomy and IFSA Collaborative Innovation Center, Shanghai Jiao Tong University, Shanghai 200240, China

[3] Department of Electrical and Computer Engineering, Old Dominion University, Norfolk, VA 23529, USA

* Email: prentzepis@tamu.edu

**Abstract**

Femtosecond, 8.04 KeV x-ray pulses are used to probe the lattice dynamics of 150 nm Cu (111) single crystal on mica substrate irradiated with 400 nm, 100 fs laser pulses. For pump fluencies below the damage and melting threshold, we observed lattice contraction due to the formation of a blast force, and coherent acoustic phonons with a period of ~69 ps. At larger pump fluence, solid to liquid phase transition, annealing, and recrystallization were measured in real time by monitoring the intensity evolution of the probing fs x-ray rocking curves and agreed with theoretical simulation results. The experimental data suggest the melting process is a purely thermal phase transition. This study provides, in real time, an ultrafast time-resolved detailed description of the significant processes that occurs as a result of a femtosecond light-pulse interacts with the Cu (111) crystal surface.

**Introduction**

The effect of ultrafast, femtosecond, pulse heating of condense matter, such as metals and semiconductors, has been associated with a number of fundamental dynamical processes that occurs on the femtosecond to picosecond temporal scales. Because the electron heat capacity is typically orders of magnitude lower than the lattice, upon femtosecond laser irradiation of a metallic sample, the photon energy is deposited by photon-electron interaction within the pulse width into the conduction band electrons located within the optical absorption depth (skin depth). The hot electrons equilibrate at a temperature of tens of thousands Kelvin within hundreds of femtoseconds by means of electron-electron interaction while the lattice system remains at room temperature. Within a few picoseconds time interval, after electron-phonon interaction, the thermal energy is transferred from the electron system to the lattice system through electron-phonon coupling until a thermal equilibrium state is established. Such a two-step non-equilibrium energy transfer process is usually described by the well-known Two-Temperature Model (TTM) [1-5]. During the first stage, the sharp temperature gradient of the hot electrons within the optical absorption depth generates a transient elastic stress, blast force[6], which develops the pressure wave, blast wave, within the laser pulse duration. Such a blast wave typically last 1~2 ps and propagates at sonic speed within the bulk of the crystal, altering the lattice parameters[7]. The properties of this blast force have been studied in detail theoretically [8-10] and its effects on the lattice has also been observed experimentally by means of time-resolved x-ray diffraction (XRD) [11, 12]. In the second stage, the energy is transferred to the lattice phonons through electron-phonon coupling within a few picoseconds, and a portion of the energy may be also carried into the bulk by ballistic hot electrons which typically have a mean free path on the order of 100 nm in metals [12]. However, in our case, the ballistic electrons generally account for a very small portion of the heat transferred, which is supported by experimental observations which show that the melting at high pump fluencies occurred only within the skin depth of 15 nm. Increase in lattice temperature causes strain and launches acoustic waves that propagate with sonic velocity within the crystal sample. For thin films grown on substrates, the blast waves and acoustic waves are reflected at the substrate, forming a standing wave between the sample surface and the substrate and causes the expansion and compression, "breathing motion", of the lattice [13]. Such coherent phonon generation has been detected in ultrafast electron diffraction studies on Al [14], and time-resolved x-ray diffraction of Au [15].

At large pump fluencies, sufficient laser energy is coupled into the lattice through photon-electron, electron-electron, electron-phonon and phonon-lattice interaction, the crystal temperature reaches the melting point and the long range lattice order disappears. The time scale of such purely thermal melting of metals is mainly determined by the electron-phonon coupling process which usually takes place within several picoseconds. Ultrafast electron diffraction studies revealed that the melting of 20nm Al and Au takes 3.5 ps and ~12 ps, respectively [16, 17], while ultrafast x-ray diffraction shown that the solid to liquid transition of thick Au crystal, 150nm, takes 8 ps [15]. In addition, non-thermal melting were also experimentally observed for semiconductors including Si, GaAs, and InSb irradiated at fluencies that are insufficient to heat them to their melting point [18-22]. The non-thermal melting is due to the laser-induced excitation of electrons that alter the interatomic potential which directly collapse the lattice. Because non-thermal melting does not involve heat transfer, it is expected to occur at a shorter time scale, namely 1 picosecond or less. However, depending on the laser fluence [23], the melting could be the combination of thermal and non-thermal mechanism and which dominates the solid to liquid phase transformation for Gold [17, 24, 25] and Aluminum [16, 26, 27] is still debated.

In this study, a 400 nm, 100 fs pulse strikes the surface of a 150 nm Cu (111) single crystal and the effect of the pulse on the crystal was monitored by femtosecond Cu K$_\alpha$ x-ray pulses. At low fluence, 10 mJ/cm$^2$, we measured, in real time, picosecond time heating and the generation of a blast force (contraction of lattice) and coherent phonons. At higher fluencies, melting, annealing, mosaic crystal formation and recrystallization was observed in the actual skin depth of the sample. The experimental data agreed with our theoretical, TTM, simulations. Monitoring these processes in real time from excitation to tens of picosecond after excitation, by means of ultrafast time-resolved XRD, we determined the dynamics and mechanism of the processes evolved, for example, the observed picosecond (~8 ps) melting suggests that melting in this case is a thermal process.

**Experimental**

The time-resolved x-ray diffraction setup has been described in detail in our previous publications [28]. The system includes a femtosecond (fs) laser system that is capable of delivering 100 mJ/pulse, 800 nm, 100 fs laser pulses at 10 Hz repetition rate, a vacuum chamber for the generation of ultrashort x-ray pulses, a sample holder that provides rotation, 3-dimentional translation, and tilting adjustments, a 2K×2K x-ray CCD with a pixel size of 13.5 μm ×13.5 μm and a linear translation stage that precisely control the relative time difference between the arrival of the pump and probe pulse on the crystal. The 800 nm fs laser pulse emitted from the fs laser system was directed into a beam splitter, with 20% of the pulse was frequency doubled to 400 nm by a BBO crystal, used to excite the 150 nm Cu (111) single crystal. The remaining 80% of the laser pulse was focused onto a 0.25-mm-diameter moving copper wire, placed in a low vacuum chamber, generating laser-plasma femtosecond Cu K$_\alpha$ x-ray pulses. A 0.3 mm slit placed 4.5 cm away from the x-ray source collimated the beam that was transmitted through the 0.1 mm Be window of the x-ray generating chamber. The single crystal Cu (111) sample was placed 14 cm away from the slit and rotated to a diffraction angle of 21.8 degrees, which corresponds to the Cu K$_\alpha$ line. The probe 8.04 KeV x-ray pulse contained $10^6$ photons/s at the sample positon. The diffraction patterns was acquired by 8 minutes integration through the x-ray CCD, that was situated approximately 21 cm from the sample. The relative delay time between the pump laser and probe x-ray pulses was precisely controlled by a linear translation stage. The 400 nm pump beam was focused onto the Cu (111) single crystal to a size of ~ 2.5 mm × 1.5 mm and a penetration depth of 15 nm. The 150 nm Cu (111) single crystal was grown on mica at a substrate temperature of 400 C. The crystal surface was measured by a scanning electron microscope and its crystal quality was determined by means of CW x-ray diffraction, which confirmed that the sample is a Cu (111) single crystal. However, such films have the usual defects that are found in epitaxial thin films. The reflectivity of Cu (111) single crystal suggested that 56% of the 400 nm pump beam energy was absorbed by the sample [29]. The diffraction of the probe x-ray pulses is governed by the first order Bragg diffraction law:

$$2d \sin \theta = \lambda \tag{1}$$

where $d$, $\theta$ and $\lambda$ are the lattice plane distance, diffraction angle and wavelength of the probing x-ray pulse, respectively. This equitation imply that, even small changes of the lattice plane distance can induce measurable shifts of the diffraction angle. Differentiating on both sides of Eq. (1) we have:

$$\frac{\Delta d}{d} = -\frac{\Delta \theta}{\tan \theta} \tag{2}$$

Therefore, the deformation of lattice plane after laser irradiation is connected with the shift of diffraction line, which is an experimentally-observable parameter. It is worth mentioning that, in ultrafast electron diffraction studies[30], the diffraction angle is typically a hundredth of a degree and $\Delta d/d \approx -\Delta\theta/\theta$ is valid due to the small angle approximation. However, this relation is invalid in time-resolved x-ray diffraction because the diffraction angle is usually large (21.8 degree for the Cu(111) sample in this paper). Typical x-ray diffraction rocking curves with and without laser excitation are shown in the insert of figure 1. The x-ray CCD array is large enough to record simultaneously the diffraction from both the excitation (signal) and non-excitation (reference) areas of the sample. Both the signal and reference areas were vertically integrated to obtain their corresponding one dimensional intensity plots and then fitted to a Gaussian function to extract the peak shift and width broadening at each delay time, which provide information regarding lattice spacing change and lattice disorders, respectively. The total intensity at each delay time is obtained by normalizing the integrated x-ray diffraction intensity of the signal area over the reference area to eliminate the pulse-to-pulse fluctuations of the probe x-ray pulse. Then, the normalized total intensities before time zero is averaged and set as the new reference to be used for the normalization of the intensities after time zero.

**Results and Discussion**

As previously discussed, upon femtosecond laser irradiation, the photon energy is initially absorbed by the free-electrons of the Cu (111) single crystal within the 15 nm penetration depth. Owing to the much smaller heat capacity of the electrons compared to the lattice system, the temperature of the electrons is initially very high (~$10^4$ K) while the crystal surface remains cold, essentially at room temperature. Then the non-equilibrium hot electrons formed within the laser penetration depth quickly transfer their energy to the electron system through electron-electron interaction and the lattice system through electron-phonon coupling, until the entire crystal reach a thermal equilibrium. The non-equilibrium hot electrons and temperature gradient contribute to the generation of blast force, sonic wave and thermal evolution at pump fluencies below the damage threshold. Increasing the photon fluency, the crystal temperature can be raised to the melting point, or above, allowing us to observe and record, in real time, the melting phase transition of the surface layer of the crystal, followed by annealing and recrystallization.

**1. Blast force, compression wave and coherent phonon generation at low laser fluence**

The peak shift of the 150 nm Cu (111) x-ray diffraction rocking curve is shown in figure 2, for pump fluencies that vary from 7 mJ/cm$^2$ to 12 mJ/cm$^2$. It is clear from Figure 2(a) that, after laser excitation, the peak shift is negative for the first few picoseconds and reach its negative maximum around 3.3 ps. After that, the peak shift moves toward the positive direction and continued to increase, followed by a damping oscillation that persisted for more than 100 ps. The negative peak shift indicate that the (111) lattice plane distance became shorted, a few ps after laser excitation, which suggests contraction of the crystal planes. This can be attributed to the theoretically predicted and experimentally observed compression wave [6, 15]. This compressive wave, due to the blast force formed initially on the surface layer of the crystal, propagates through the bulk of the crystal, therefore, the surface layer of the sample is initially denser with a reduced lattice plane distance. As a consequence, the experimental data shown in figure 2(a) reveals a negative peak shift for the first few picoseconds. The contraction between such excitation areas and the cold areas of the crystal occurs before phonon-phonon interaction and thermalization. Because the laser pulse penetrates only through the few top lattice planes of the sample (15 nm), while the x-ray pulse probes the entire 150 nm thickness of the sample, it is expected that the x-ray contraction signal observed accounts for only ~10% of the overall diffraction signal. As can be seen in figure 2(a), the contraction represented by the shift signal is, proportionally small, around $2\times10^{-4}$. After the contraction stage, the electron-phonon and the phonon-phonon coupling take place and contribute to the establishment of the thermal equilibrium state of the entire crystal. Therefore, the thermal energy deposited into the skin depth propagates through the bulk and heats the entire depth of the crystal. The elevated temperature of the crystal induces thermal expansion which was displayed by the increased peak shift of the x-ray diffraction rocking curve, illustrated in figure 2. In addition to the increased peak shift, several damping oscillations are also observed in the current experiments. Those periodic oscillation, that represents lattice vibration, indicate generation of coherent phonons [31]. Similar lattice vibrations have been observed in time-resolved electron diffraction studies of 20 nm thin Al crystal [32, 33]. Previous time-resolved XRD experiments on 400 nm Ge single crystal also revealed similar damping oscillation of the lattice planes [34]. Theoretical studies based on the two-temperature model (TTM) and the Fermi-Pasta-Ulam anharmonic chain model have been applied to explain the experimentally observed acoustic oscillations [35, 36]. It has been shown in previous time-resolved XRD studies of Au single crystal that, the laser fluence does not affect, significantly the period of the coherent phonon oscillations [12]. Similar results are found in the present study for the copper crystals as shown in Figure 2(b) for two different pump fluencies of 7 mJ/cm$^2$ and 12 mJ/cm$^2$. The oscillations can be simplified to a one-dimensional standing wave between the crystal surface and mica substrate. The oscillation period, T, may be calculated using the longitudinal velocity of the acoustic wave, $v$ in solid copper:

$$T = 2L/v \qquad (3)$$

Where L=150 nm is the thickness of the copper single crystal sample and 4660 m/s the sound velocity in solid copper. Therefore, the calculated oscillation period T is 64 ps, which agrees with our observation of ~ 6 ps, average of the two damping periods shown in figure 2. The damping of those oscillation is attributed to the energy dissipated, mainly, onto the mica substrate. The damping takes about 180 ps and after that the peak shift of the x-ray rocking curve remains unchanged up to the longest delay time, 200 ps, in our experiments.

The temperature change in the lattice is associated with the change in lattice plane distance $\Delta d$, and therefore the peak shift of x-ray rocking curve. Designating the thermal expansion coefficient of copper by $\alpha$, it can be easily shown that:

$$\Delta T_l = \frac{1}{\alpha} \frac{\Delta d}{d} = -\frac{1}{\alpha} \frac{\Delta \theta}{\tan \theta} \qquad (4)$$

Assuming a purely thermal process, the temperature change of lattice $\Delta T_l$ is linearly depended on the laser intensity absorbed by the sample $I_{abs}$. We have:

$$\Delta \theta \propto \Delta T_l \propto I_{abs} \qquad (5)$$

Increasing the pump laser fluency, it is expected that the peak shift of x-ray rocking curve would increase linearly. However, it should be noted that temperature is defined as a statistic parameter that describes the equilibrium state of a large number of particles. Given the non-equilibrium heating of the crystal due to the ultrafast energy deposition through femtosecond laser pulse excitation, the lattice temperature being discussed here is the "equivalent temperature" which assumes that temperatures can be defined around each lattice position at a particular time. This is generally used definition of temperature in ultrashort laser pulse-matter interaction studies. This equivalent temperature takes each unit cell as a mini equilibrium system with representative electron and lattice temperatures and uses them to define the temperature of the entire crystal, which is a non-equilibrium system. We increased the pump laser fluency, from 3.5 mJ/cm$^2$ to 17

mJ/cm$^2$, and plotted the peak shift in figure 3. The two representative peak shifts of each pump fluency are chosen at delay times of 35 ps and 180 ps. At a delay time of 35 ps, the oscillations reach their maximum amplitude while at around 180 ps delay time, the peak shift is almost at a plateau. The linear relation indicated by the log-log plot of figure 3 agrees well with the theoretical analysis of Eq. (3) and also shows that the Cu single crystal heated by ultrashort laser pulse is a thermal process.

The broadening of the XRD rocking curve vs delay times at two different pump energies are shown in Figure 4(a) and the pump fluency vs broadening at two representative delay times, 35 ps and 180 ps, is depicted in Figure 4(b). The oscillation period of the broadening is the same as that observed in the peak shift, indicating the propagation of sonic waves between the surface layer of the crystal and the substrate. According to previous studies, the broadening due to the blast force and thermal stress depends upon the second order of pump laser intensity and lattice temperature [8-12, 37]. Therefore, the relation between the broadening of the rocking curve, $\Delta FWHM$ and the pump laser intensity, $I_{abs}$ is given by:

$$\Delta FWHM \propto \left(\Delta T_l\right)^2 \propto I_{abs}^2 \qquad (6)$$

As depicted in Figure 4(b), the linear fitting on the log-log plot shows that the broadening at 35ps and 180ps change with slopes of 1.5 and 2.1, respectively. Those slopes agree with the prediction of Eq. 6, which further provides evidence that the broadening caused by the non-thermal expansion and compression of the x-ray rocking curve is due to the stress and tension generated from the blast and sonic waves within the crystal.

**2. Melting and annealing of the Cu (111) single crystal at high laser fluence**

With the excitation fluency gradually increased to the melting temperature of copper, 1357K, solid-to-liquid phase transformation is expected to occur on the 15 nm skin depth of the crystal, followed by annealing, mosaic crystal formation, and recrystallization. Time-resolved electron diffraction, X-ray scattering, and optical spectroscopic techniques have been previously utilized to study melting of metals and semiconductors [38-44]. In this study, we observed the solid-liquid and liquid-solid phase transformation and other processes of Cu (111) single crystal induced by irradiating it with intense 400 nm femtosecond laser pulses. The crystal structure was monitored through the temporal evolution of the XRD intensity. The pump fluency used, ~58 mJ/cm$^2$, was insufficient to induce damage to the crystal while intense and energetic enough to trigger melting within the skin depth. As depicted in Figure 5, at negative delay times, before excitation, the XRD signal is in all aspects the same as those recorded at room temperature, without laser irradiation. At delay times of ~8 ps after laser irradiation, a sharp decrease of the XRD intensity of the 150 nm Cu (111) single crystal was observed. After that, the XRD signal intensity begin to increase until ~ 40 ps and then it takes more than 80 ps for the diffraction intensity to recover to its original value. Diffraction signals recorded after the crystal is cooled down to room temperature are the same as those obtained at the start of the experiment, without fs pulse excitation, which indicates that the copper crystal remains undamaged throughout the experiment. The entire process is attributed to the melting and annealing of Cu (111) single crystal which generally consist of three steps: (i) the removal of crystal defects that cause the internal stresses inside the crystal by means of melting or softening; (ii) the growing of grains that introduces new internal stress if the crystal temperature remains high enough to maintain annealing condition; (iii) the mosaic crystal formation and recrystallization process that removes the internal stresses owing to nucleation and growing of new strain-free grains.

The decrease of the XRD intensity, ~7%, during the first few picoseconds is attributed to melting or atomic disorder of the Cu (111) single crystal. Due to the limited fs light absorption depth, the picosecond disorder process occurs only within the optical skin depth (15nm for 400nm pump laser wavelength). However, the femtosecond x-ray pulses penetrate through and probe the entire 150 nm thick copper crystal. As a consequence, only a small portion of approximately 10%, 15 nm skin depth, is expected to be affected by melting. This also suggest that ballistic electrons are few and do not affect the melting process. If a significant amount of the laser energy is carried into the bulk of the sample by ballistic electrons, the entire sample would melt within the first few picosecond and the total diffraction intensity will decrease by nearly 100%, which means no diffraction signal would be observed. The experimental data presented here also show that the rate of melting is estimated to be $1.5 \times 10^{11}$ $s^{-1}$. Given that the electron-phonon interaction and phonon-lattice interaction are on the order of picoseconds, our experimental data suggest that thermal melting, rather than interatomic potential change due to electronic excitation, is the main mechanism, this agrees with previously published reports [15].

The increase of the XRD intensity from ~93%, at about 8ps, to approximately 114%, at 40ps, after laser excitation is attributed to softening, mosaic crystal formation and recrystallization. For the laser fluence used for melting within surface layer, we expected that only softening of interatomic potential of the inner area of the crystal will occur [45]. Considering the heat capacity of energetic electrons is on the order of $10^6$ $J/m^{-3}K$, the electron temperature within the skin depth is estimated to be $10^4$ K during the laser irradiation period. Those hot electrons quickly equilibrated and transferred energy into the cold lattice of the skin depth (surface area) by electron-phonon coupling and cause, within picoseconds, the disappearing of long range order, in that area. Meanwhile, the surface area that is at elevated temperatures, also transfers energy into the bulk of the Cu (111) crystal through thermal diffusion. As the heat energy transfers from the skin to the inner area of the crystal through phonon-phonon interaction, the temperature of the surface layer decreases. When the temperature of the surface layer decreases below the melting point, recrystallization occurs. Meanwhile, the heat transferred to the inner crystal area is expected to induce softening. This dynamics of the processes involves both the interior deformation of the crystal and the crystallization of the skin. Grains are microscopic crystals held together through their boundaries where crystals of different orientations meet. During the nucleation, which results from the decreased temperature, the size of the grains increase and the boundaries between grains decrease in number. With the growing of grains, blocks of mosaic crystal are formed with a dimension on the order of $10^{-5}$ cm which are tilted by fractions of a minute of arc with respect to one another. Because the x-ray beam is not ideally collimated but has a small divergence angle of 0.4 degree, more x-rays within the illumination area will be diffracted by the slightly tilted mosaic crystals. Due to the incoherence of diffraction waves from different mosaic crystal areas, the total XRD intensity is the sum of individual signals from each mosaic block. Therefore, the observed XRD intensity, at around tens of picoseconds, may exceed the diffraction intensity of original crystal. During anneal of the crystal, the mosaic crystal "fuse" into a single-orientation (single crystal), which is monitored by the decrease of XRD intensity toward their value before excitation. The effects on diffraction intensities owing to mosaic crystals formation have been theoretically pioneered by C. G Darwin in earlier 1920s [46-48] and have been experimentally observed [49, 50]. It is also worthy to note that the dip of the diffraction peak position change due to the compression wave occurred around 3.3 ps while the melting revealed by the intensity drop is around 8 ps, which provides experimental evidence that the strain revealed by the peak breath is unrelated to the intensity drop caused by lattice collapse. Similar observations have also been reported in ultrafast electron diffraction experiments [33, 51].

The laser heating of the Cu (111) crystal is simulated through the two-temperature model (TTM) and the results are shown in Figure 6. Because the laser irradiation area is much larger than signal area probed by the x-ray beam, the three-dimensional laser-metal interaction is reduced to one dimension. Therefore, the electron and lattice temperature at a given depth below the sample surface for a particular delay time can be described by $T_e(z,t)$ and $T_l(z,t)$, respectively. According to TTM and considering the irradiation by the femtosecond laser pulse $S(z,t)$, the evolution of electron and lattice temperatures are described by two coupled equations:

$$C_e \frac{\partial}{\partial t}T_e(z,t) = \frac{\partial}{\partial z}\left(\kappa_e \frac{\partial}{\partial z}T_e(z,t)\right) - g\left[T_e(z,t) - T_L(z,t)\right] + S(z,t) \quad (7)$$

$$C_L \frac{\partial}{\partial t}T_L(z,t) = g\left[T_e(z,t) - T_L(z,t)\right] \quad (8)$$

At low pump fluencies where the electron temperature is below thousands of Kelvin, it is a valid approximation to assume that the electron-phonon coupling constant, $g$, is a constant; the electron specific heat capacity $C_e$ is linearly dependent on electron temperature; and the thermal conductivity of electrons, $\kappa_e$, is also linearly dependent on the ratio of electron and lattice temperature[52]. However, for the high laser fluence used for melting in current experiment, the electron temperature reaches more than $10^4$ K. Therefore, the simple temperature dependence of $g$, $C_e$ and $\kappa_e$ are invalid and more accurate modeling of those parameters are required. In our simulation, taking into account the effects of electron-electron, and electron-phonon scattering on the electron relaxation time, the electronic thermal conductivity based on Drude model is used [53-55]:

$$\kappa_e = \frac{a_0 T_e}{a_1 T_e^2 + a_2 T_L} \tag{9}$$

in which $a_0 = 9.7 \times 10^{13} Wm^{-1}K^{-2}s^{-1}$, $a_1 = 2.66 \times 10^6 K^{-2}s^{-1}$, and $a_2 = 2.41 \times 10^{11} K^{-1}s^{-1}$. The electron-phonon coupling, $g$, and electron heat capacity, $C_e$, are obtained through fifth order Padé approximations of the data presented in [55] and [52], respectively:

$$g = 1 \times 10^{17} \frac{\sum_{n=0}^{5} A_1(n) \left(\frac{T_e}{10^4}\right)^n}{1 + \sum_{n=1}^{5} A_2(n) \left(\frac{T_e}{10^4}\right)^n} \tag{10}$$

$$C_e = 1 \times 10^5 \frac{\sum_{n=0}^{5} B_1(n) \left(\frac{T_e}{10^4}\right)^n}{1 + \sum_{n=1}^{5} B_2(n) \left(\frac{T_e}{10^4}\right)^n} \tag{11}$$

where $A_1(0) = 0.561$, $A_1(1) = -2.263$, $A_1(2) = 9.436$, $A_1(3) = -32.906$, $A_1(4) = 64.683$ and $A_1(5) = 40.393$; $A_2(1) = -3.587$, $A_2(2) = 9.103$, $A_2(3) = -6.258$, $A_2(4) = 8.118$ and $A_2(5) = 6.725$; $B_1(0) = -0.039$, $B_1(1) = 11.995$, $B_1(2) = -76.675$, $B_1(3) = 222.761$, $B_1(4) = -238.574$ and $B_1(5) = 175.460$; $B_2(1) = -5.700$, $B_2(2) = 15.707$, $B_2(3) = -19.342$, $B_2(4) = 13.361$ and $B_2(5) = -0.865$. The TTM equations (7) and (8) were solved using an implicit finite-difference scheme with a grid spacing equal to the lattice plane distance of Cu (111) single crystal. During the simulation, the heat capacity of Cu lattice, $C_L$, was taken as $3.445 \times 10^6 Jm^{-3}K^{-1}$. And at room temperature, the reduced values for $g$, $C_e$ and $\kappa_e$ are $5.55 \times 10^{16} Wm^{-3}K^{-1}$, $105 Jm^{-3}K^{-2}$ and $401 Wm^{-1}K^{-1}$.

The two-dimensional plot of the spatial-temporal evolution of electron and lattice temperature is given by Figure 6, where the electron temperature is shown to reach its highest value, $\sim 10^4$ K, within 1 ps after laser irradiation, followed by a thermalization within 2 ps. Accompanying the heat transfer from the electron to the lattice, the phonon temperature reaches the melting point in approximately 5 ps. The melting of the surface layer reached the maximum melting depth of 9 nm in around 10 ps. Those results agree with our experimental data which show that the diffraction intensity dropped by 7% within 8 ps, indicating that about 10 nm of the surface layer has melted during this time.

**Conclusion**

Direct measurements of transient structure changes in Cu (111) single crystal illuminated with 400 nm, 100 fs laser pulses have been obtained by means of ultrafast fs time-resolved x-ray diffraction. Monitoring the picosecond evolution of X-ray diffraction intensity, width, and shift, we have determined directly, in real time, the electron-phonon coupling, phonon lattice interaction, and the evolution of lattice order, such as the lattice compression due to blast force, breathing motion due to the propagating of acoustic waves, melting, nucleation, mosaic crystal formation and recrystallization. The long range lattice order disappears within 8 picoseconds, which suggests a thermal melting mechanism. Theoretical simulations, TTM, of the femtosecond pulse interaction with the Cu (111) crystal including electron temperature, lattice temperature, and melting agree well with the experimental data presented.

**Acknowledgements**

This research was supported by The Welch Foundation Grant Number: 1501928, Texas A&M University TEES funds, Texas A&M University at Qatar, and The National Natural Science Foundation of China Grant Number: 61222509.

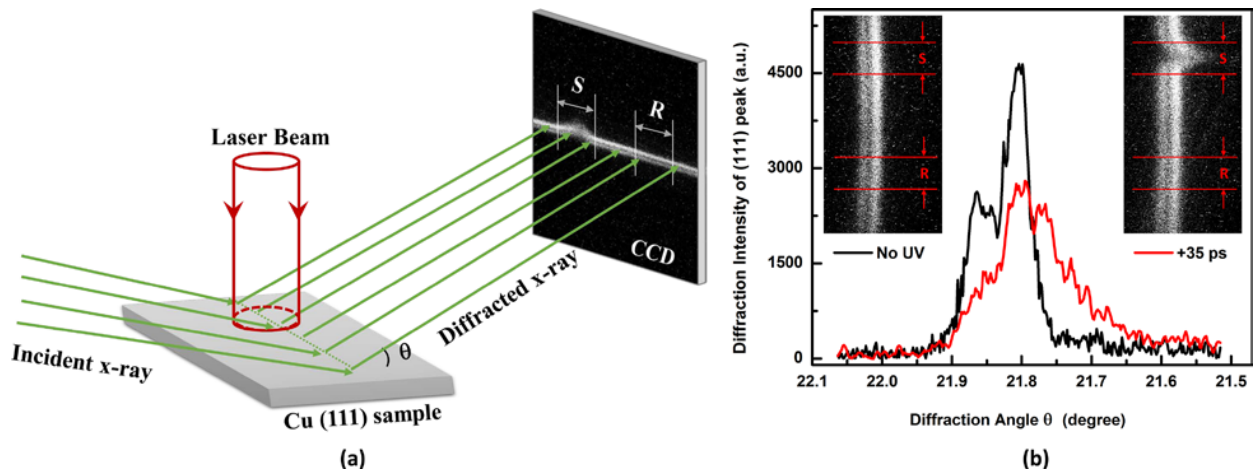

Figure 1: (a) Geometric relationship of the sample, the detector, probing x-rays and pumping laser. (b)Rocking curve of the heated and cold areas of the 150 nm Cu (111) sample, at 35 ps delay time and before laser irradiation. The pump fluency is 17.5 mJ/cm$^2$. Insert: CCD image recorded after 8 min exposure. Diffraction signals from the excited and non-excited areas of the sample are marked as S and R, respectively. The double lines of the diffraction pattern are the k$\alpha_1$ and k$\alpha_2$ diffractions.

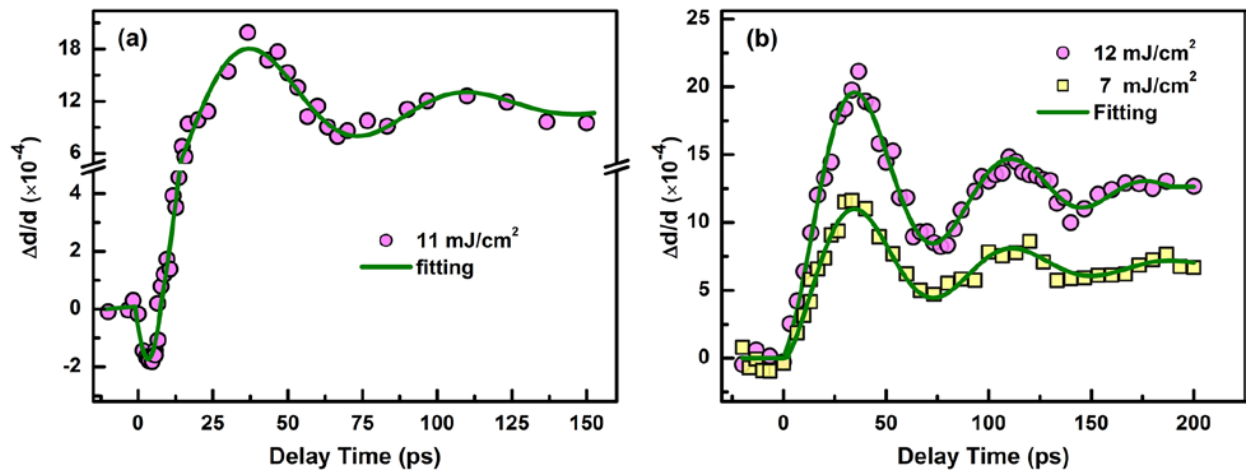

Figure 2: Peak center shift of the x-ray diffraction rocking curve. (a) Peak shift at a pump fluence of 11 mJ/cm$^2$. The contraction of lattice around the first few picoseconds is due to the blast wave. (b) Peak shift at pump fluencies of 7 mJ/cm$^2$ and 12 mJ/cm$^2$. The contraction is not observed because of the large delay time step, ~14ps, used during the data acquisition.

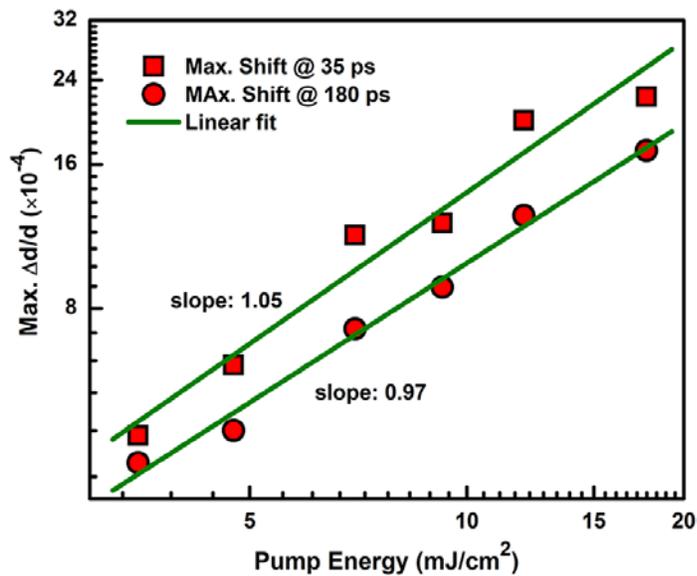

Figure 3: Energy dependence of the peak shift at two representative delay times, namely 35 ps and 180 ps. The linear fitting in the log-log plot indicates a linear slope.

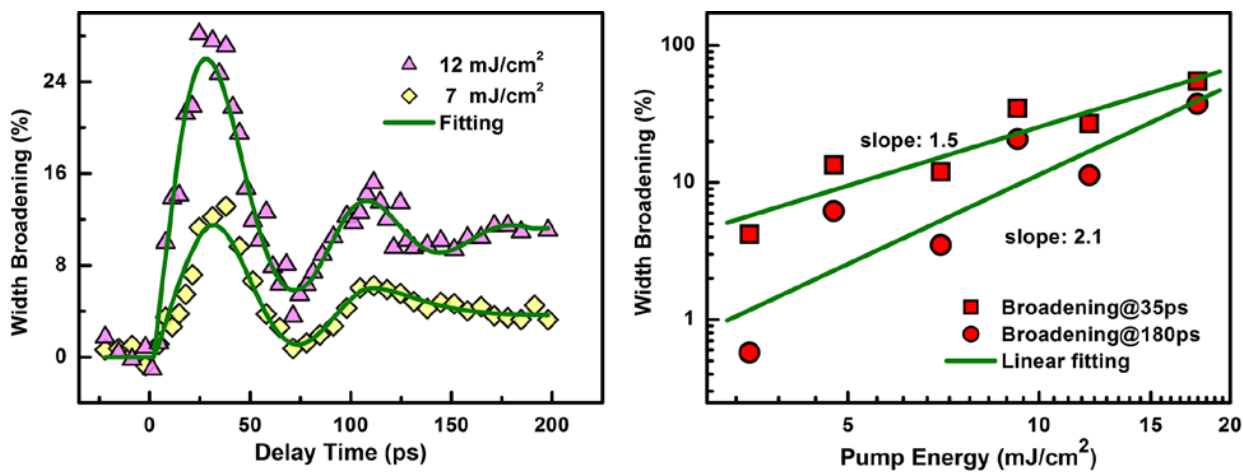

Figure 4: (a) Diffraction line broadening at pump fluencies 7 mJ/cm$^2$ and 12 mJ/cm$^2$; (b) Pump fluency dependent broadening of the XRD rocking curve at two representative delay times: 35ps and 180 ps. The linear fit in the log-log plot shows slopes of 1.5 and 2, which indicate a nonlinear dependence of broadening on pump fluency.

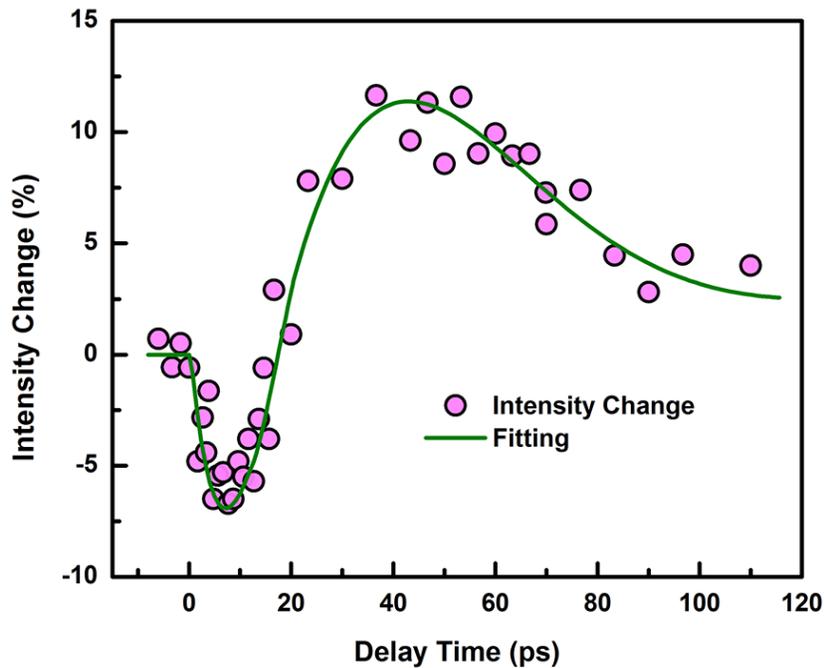

Figure 5: The time-dependent intensity change of the x-ray rocking curve. The 150 nm Cu (111) single crystal is irradiated with 400 nm, 100 fs, ~58 mJ/cm$^2$ laser pulse.

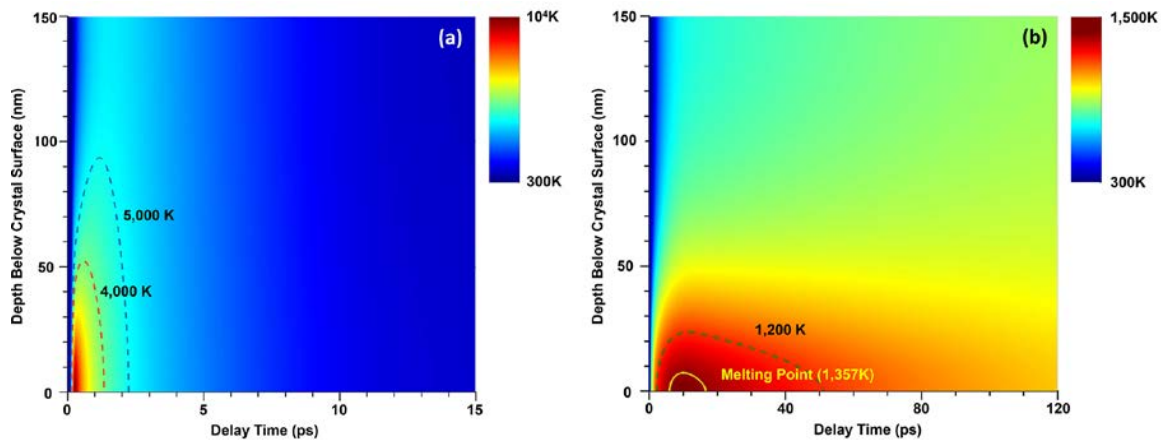

Figure 6: Spatial-temporal distribution of (a) electron and (b) lattice temperature of Cu (111) single crystal simulated by two-temperature model. The contour curve in (b) indicated the melting point of Cu.